\documentclass{jfm}
\usepackage{graphicx}
\usepackage{epstopdf, epsfig}
\usepackage{comment}
\usepackage{amsmath}
\usepackage[dvipsnames]{xcolor}

\shorttitle{Finite-size bubbles in Taylor-Couette turbulence}
\shortauthor{V. Spandan, R. Verzicco and D. Lohse}

\title{Physical mechanisms governing drag reduction in turbulent Taylor-Couette flow with finite-size deformable bubbles}

\author[Vamsi Spandan, Roberto Verzicco, Detlef Lohse]%
{Vamsi Spandan$^1$,\ns Roberto Verzicco$^{2,1}$, and Detlef Lohse$^{1,3}$%
\thanks{Email address for correspondence: d.lohse@utwente.nl }}

\affiliation{$^1$Physics of Fluids Group and Max-Planck , Faculty of Science and Technology, J.M. Burgers Center for Fluid Dynamics and MESA+ Institute, University of Twente, 7500 AE Enschede, Netherlands\\[\affilskip]
$^2$Dipartimento di Ingegneria Meccanica,University of Rome \lq Tor Vergata\rq, Via del Politecnico 1,
Rome 00133, Italy\\[\affilskip]
$^3$Max Planck Institute for Dynamics and Self-Organization, Am Fassberg, 37077 G\"ottingen, Germany
}

\begin{document}

\maketitle

\begin{abstract}
The phenomenon of drag reduction induced by injection of bubbles into a turbulent carrier fluid has been known for a long time; the governing control parameters and underlying physics is however not well understood. In this paper, we use three dimensional numerical simulations to uncover the effect of deformability of bubbles injected in a turbulent Taylor-Couette flow on the overall drag experienced by the system. We consider two different Reynolds numbers for the carrier flow, i.e. $Re_i=5\times 10^3$ and $Re_i=2\times 10^4$; the deformability of the bubbles is controlled through the Weber number which is varied in the range $We=0.01 - 2.0$. Our numerical simulations show that increasing the deformability of bubbles i.e., $We$ leads to an increase in drag reduction. We look at the different physical effects contributing to drag reduction and analyse their individual contributions with increasing bubble deformability. Profiles of local angular velocity flux show that in the presence of bubbles, turbulence is enhanced near the inner cylinder while attenuated in the bulk and near the outer cylinder. We connect the increase in drag reduction to the decrease in dissipation in the wake of highly deformed bubbles near the inner cylinder.  
\end{abstract}


\section{Introduction}
\label{sec:intro}

Drag in turbulent flows is a major drain of energy. It is well known that the addition of a secondary phase such as polymers or bubbles into a turbulent carrier fluid can lead to significant reduction in the overall drag in the system \citep{van2005drag,van2013importance,verschoof2016bubble,white2008mechanics,ceccio2010friction}. Here, we define drag reduction ($DR$) as 
\begin{equation}
DR [\%] = 100\times\frac{C_{f,0}-C_{f,\alpha}}{C_{f,0}}
\label{eqn:cf}
\end{equation}
where $C_{f,0}$ and $C_{f,\alpha}$ are friction coefficients with $\alpha$ volume fraction of the secondary phase injected into the system. For example, a small volume fraction of bubbles ($<$ 4\%) injected into a turbulent carrier liquid can reduce the overall drag of the system by up to 40\%  \citep{van2013importance,verschoof2016bubble}. The extent of modification in momentum transport that can be achieved through bubble injection makes the phenomena relevant for both fundamental studies of multiphase flows and large-scale industrial applications \citep{ceccio2010friction}. However, given the many challenges and large number of control parameters in theoretical, experimental and numerical techniques, the governing physics in multiphase flows is not fully understood. The key parameters that play a role in the overall drag reduction measured in a multiphase turbulent system are the Reynolds number of the carrier fluid, the geometry of the flow, the relative size of the dispersed phase in comparison to the Kolmogorov scale (either sub-Kolmogorov or finite-size), gravity, volume fraction and surface-tension of the dispersed phase. In this paper, we use Direct Numerical Simulations (DNS) to uncover the effect of deformability of finite-size bubbles on the overall drag in a canonical wall-bounded shear driven system, namely in Taylor-Couette (TC) flow.  

In TC flow, the fluid is driven by two independently rotating co-axial cylinders. It has been one of the classical systems to study shear driven fluid dynamics for the past several decades (see \citet{grossman2016high} for a recent review). The geometry of TC flow is closed, simple and allows for exact global balance relations between the driving and dissipation in the flow, which opens up the possibility for high-precision experiments and numerical simulations. In the context of multiphase TC flow, experimental and numerical studies have demonstrated that the overall drag in the system can be reduced with the injection of a small volume fraction of bubbles \citep{murai2005bubble,murai2008frictional,sugiyama2008microbubbly,spandan2016drag}. Given the simple and closed nature of TC flow, in comparison to other canonical systems, multiphase TC flow is an ideal playground to understand the governing physics behind bubble-induced drag reduction. In contrast, studies on bubble induced drag reduction in classical plane channel flows suffers from limitations in the domain-size in the stream-wise direction. TC flow on the other hand is naturally limited in the stream-wise (azimuthal) direction, thus minimising any domain-size effects. 
%Additionally, multiphase TC flow allows for an in-depth study into the interaction between boundary layers, bulk and the immersed bubbles.  

Various mechanisms have been proposed to explain the origin of bubble induced drag reduction effect \citep{ferrante2004physical,lu2005effect, lu2008effect, l2005drag, sugiyama2008microbubbly, van2007bubbly, ceccio2010friction, van2013importance, verschoof2016bubble, spandan2016drag, spandan2017deformable}. When the carrier fluid is weakly turbulent (laminar boundary layers and turbulent bulk), the effective buoyancy of dispersed sub-Kolmogorov bubbles, which is described by the Froude number, plays an important role in governing drag reduction in the system \citep{spandan2016drag}. When the carrier fluid becomes highly turbulent (both boundary layer and bulk are turbulent), the buoyancy of sub-Kolmogorov bubbles is not sufficient any more to achieve strong drag reduction effects. Numerical simulations have shown that in the highly turbulent regime, even deformability of sub-Kolmogorov bubbles does not contribute to drag reduction \citep{spandan2017deformable}. In the highly turbulent regime, deformability of finite-size (larger than Kolmogorov scale) bubbles seems to be a promising mechanism to achieve strong drag reduction \citep{van2013importance,verschoof2016bubble}. However, it is unclear why the extent of deformability of finite-size bubbles is important for achieving strong drag reduction in highly turbulent TC flow. 

In this paper, we show that the origin of deformability induced drag reduction is linked to reduced dissipation in the wake of finite-size (larger than Kolmogorov scale) deformable bubbles in comparison to rigid spherical bubbles. Using fully resolved numerical simulations we are able to identify and separate the various effects contributing to drag reduction in the flow and furthermore look at the individual contributions from these effects. We find that the reduced dissipation in the bubble wake is the dominant mechanism of drag reduction irrespective of whether the carrier fluid is weakly or highly turbulent. 

\section{Numerical Details and Parameters}
\label{sec:num}

The dynamics of the carrier phase is solved using DNS of the Navier-Stokes equations in cylindrical coordinates. The governing equations for the carrier phase are

\begin{equation}
\frac{\partial \textbf u}{\partial t}+\textbf u \cdot \nabla \textbf u=-\nabla p+\frac{1}{Re}\nabla^2 \textbf u +\textbf f_b(\textbf x,t),
\label{eqn:ns}
\end{equation}
\begin{equation}
\nabla \cdot \textbf u=0 .
\label{eqn:con}
\end{equation}
$\textbf u$ and $p$ are the carrier phase velocity and pressure, respectively, while $\textbf f_b(\textbf x,t)$ is a source term included in the fluid momentum equation arising from the immersed boundary method (IBM) and is used to enforce the interfacial boundary condition at the bubble-fluid interface. A second-order accurate finite-difference scheme with fractional time stepping is used for the spatial and temporal discretisation of equations (\ref{eqn:ns}) and (\ref{eqn:con}) \citep{verzicco1996finite,van2015pencil}. Unlike sub-Kolmogorov bubbles which can be modelled using the point-particle approximation \citep{elghobashi1994predicting}, finite-size bubbles which are much bigger than the Kolmogorov scale experience non-uniform flow conditions on the surface. In such a case, momentum exchange between the carrier fluid and dispersed bubbles cannot be modelled using point-wise approximations, but has to be fully resolved. This is done using IBM where each dispersed bubble is discretised using an unstructured Lagrangian mesh which resides over a structured Eulerian mesh on which equations (\ref{eqn:ns}) and (\ref{eqn:con}) are solved. The influence of the dispersed bubbles onto the carrier fluid is accounted for through a volume averaged force first computed on the Lagrangian mesh and then transferred to the Eulerian mesh in a conservative manner. 
A no-slip boundary condition is imposed on the interface of each bubble which physically corresponds to a contaminated interface. While the IBM also allows for imposing a free-slip boundary condition which replicates a free-slip interface, we restrict ourselves to contaminated interfaces in this work as it can be handled easily from a numerical point of view. Due to the lack of a boundary layer along the surface of clean bubbles with a free-slip interface, it can be expected that the resulting local dissipation is lower in comparison to contaminated no-slip bubbles. This may lead to an increase in the magnitude of drag reduction in comparison to contaminated no-slip bubbles. In this study, we focus on understanding the physical mechanisms relevant for increasing drag reduction with increasing deformability of contaminated no-slip finite-size bubbles. 

The deformation dynamics of the immersed bubbles is computed through an interaction potential (IP) approach where the characteristic surface tension of a liquid-liquid interface is replicated using a triangulated network of in-plane elastic and out-of-plane torsional springs. The IP approach for modelling deformation of immersed liquid-liquid interfaces has been proven to be reliable and self-consistent as long as the immersed interfaces do not approach their breakup limit. The basic idea behind the IP approach is that under the action of external forces, the immersed interface adjusts itself through the action of internal spring forces such that the total displacement potential energy tends to a minimum. All bubbles are initialised with a spherical triangulated network of springs. While external forces such as local pressure fluctuations and viscous stresses deform the surface, the internal spring forces tends to bring the deformed surface back to its initial spherical shape with a goal to minimise the overall potential energy stored in the network. Given the finite-size nature of the immersed bubbles, the pressure fluctuations and viscous stresses are computed locally on individual triangular elements \citep{spandan2017parallel}. The elastic constants required to model the triangulated network in the IP approach for interfaces is chosen through a tuning procedure. The tuning of the surface properties of the triangulated network and its corresponding Weber number (defined later) is performed for a centroid fixed bubble in a cross flow as described in \citet{spandan2017parallel}. Here, it is important to note that the IP model is not an exact representation of the surface tension phenomenon and it is to be taken as a phenomenological approach to mimic the deformation characteristic of immersed drops or bubbles. Additional details on the IBM, the interaction potential approach and the parallelisation schemes can be found in \citet{spandan2017parallel} and \citet{detullio2016moving}.

The geometrical control parameters in TC flow are the radius ratio $\eta=r_i/r_o$ and the aspect ratio $\Gamma=L/(r_o-r_i)$, where $r_i$, $r_o$ are the radii of the inner cylinder (IC) and outer cylinder (OC), respectively and $L$ is the height of the cylinders. In this paper, we set $\eta=0.909$ and $\Gamma=2.0$. The driving in the system is characterised by the IC Reynolds number $Re_i=Ud/\nu$, where $U$ is the velocity of the IC, $d=r_o-r_i$ is the gap-width between the cylinders and $\nu$ is the kinematic viscosity of the carrier fluid; the OC is kept stationary. In the following section, we study the effect of finite-size deformable bubbles on the carrier fluid dynamics for two different IC Reynolds numbers i.e. $Re_i=5\times 10^3$ and $Re_i=2\times 10^4$, which are hereafter referred to as low and high $Re_i$ cases, respectively. The extent of deformability of the immersed bubbles is controlled through the Weber number $We=\rho_fU^2d_p/\sigma$, where $\rho_f$, $d_p$ and $\sigma$ are the fluid density, bubble diameter and surface tension, respectively. 
The Weber number of the bubbles considered are $We=10^{-2}, 0.5, 1.0, 2.0$. In order to simulate rigid non-deformable bubbles ($We=10^{-2}$), numerical stiffness is avoided by treating the immersed bubbles as rigid spheres rather than compute the small scale deformations from the IP model. The non-dimensional in-plane elastic constants for $We=0.5, 1.0, 2.0$ are $k_e^*=12.0\times 10^{-3}, 8\times 10^{-3}, 5\times 10^{-3}$. The remaining constants used for the triangulated network are scaled according to the guidelines specified in \citet{spandan2017parallel}; i.e. bending constant $k_b^*=k_e/10.0$, volume constant $k_v^* \sim 10^5 k_e^*$ and area constant $k_a^*=k_e^*$. The ratio between the mean edge length on the initial triangulated sphere to the gap-width between the cylinder is approximately $7.5\times 10^{-3}$.
%where $We=10^{-2}$ and $We=2.0$ corresponds to rigid non-deformable and highly deformable bubbles, respectively. 

In this paper, the main focus is to understand the effect of Weber number $We$ which quantifies the ratio between inertial and surface tension forces acting on the immersed bubbles. 
The choice of the reference velocity in the definition of $We$ varies from one study to the other; 
%for example in previous experiments \citep{van2013importance,verschoof2016bubble} the reference velocity is taken to be the liquid velocity fluctuation at the position of the bubble. 
regardless of the definitions in simulations and experiments, the main idea is that the Weber number indicates the extent of deformability of the dispersed phase.
%; i.e. high value of $We$ implies a deformable bubble/drop while low $We$ implies the bubble/drop is rigid and difficult to deform. 
We only consider bubbles which do not breakup or coalesce in the flow; modelling the breakup and coalescence dynamics of bubbles in a turbulent environment is an extremely challenging task from both a physical and numerical point of view and is not in the scope of this study. Collision among the immersed bubbles and of the bubbles with the cylinder walls is implemented through an elastic potential between the Lagrangian markers and the centre of the Eulerian cell in which the Lagrangian marker resides. The numerical algorithm is as follows. At every time-step, when two or more Lagrangian markers from different bubbles reside in the same Eulerian cell, all the Lagrangian markers in that Eulerian cell face a repulsive force proportional to the square of the inverse distance between the marker and the centre of the Eulerian cell. This ensures that different bubbles never come into contact with each other thus avoiding any \lq ghost\rq\ overlap between the Lagrangian meshes \citep{spandan2018fast}. In the following simulations, in addition to ensuring energy conservation, we keep the same formulation of the collision elastic potential in all cases to eliminate any numerical artefacts among the different cases. The grid resolution for the carrier fluid is set to $N_\theta \times N_r \times N_z = 768\times 192\times 384$ and $N_\theta \times N_r \times N_z = 1536\times 384\times 768$ for the low and high $Re_i$ cases, respectively. 

A total of 120 bubbles are simulated along with a turbulent carrier fluid which corresponds to a global volume fraction of 0.1\%; each individual bubble is discretised using 1280 and 2560 Lagrangian markers for the low and high $Re_i$ cases, respectively. The density ratio of the bubbles to that of the carrier fluid is taken to be approximately $\tilde \rho=\rho_b/\rho_f=0.05$, which is larger than real gas bubbles but easier to handle numerically. 
The effective Froude number of the bubbles in all the simulations is kept the same and is defined as $Fr=\sqrt{\tilde \rho U^2/(\tilde \rho-1)gr_i)}=0.64$. For the IBM, we assume that the viscous forces acting on the interface from the fluid entrapped inside the immersed \lq bubble\rq\ is close to negligible; in the context of the IBM, this would correspond to a ratio of bubble-gas viscosity and carrier fluid to be approximately zero. 
To achieve numerical convergence, we run our simulations in a series of three steps. First, we allow the single phase system to converge. We then initialise rigid bubbles and allow them to develop in the flow. Next, we allow them to deform corresponding to their imposed Weber number and run the simulation until the two-phase system has the developed statistical stationarity. The convergence criteria is described later. Since we keep the relative diameter ($d_p$) of the bubbles  with respect to the gap width between the cylinders the same for both the low and high $Re_i$ cases, the relative size of the bubbles in comparison to the Kolmogorov scale ($\eta_K$) of the corresponding single phase flow is approximately $d_p/\eta_K\sim 14$ and $d_p/\eta_K\sim 25$ for the low and high $Re_i$ case, respectively.

\section{Results}
\label{sec:res}

\begin{figure}
\centering
\includegraphics[scale=0.75]{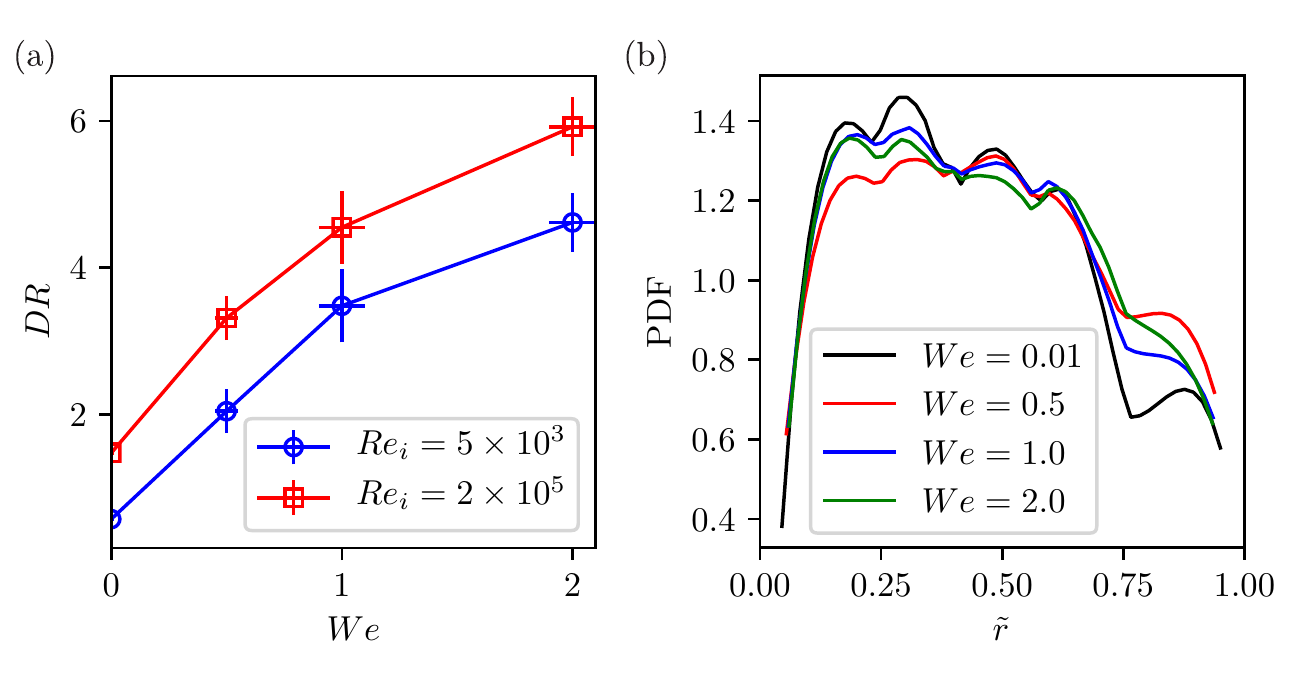}
\caption{(a) Drag reduction ($DR$) versus Weber ($We$) number. (b) Probability distribution function (PDF) of radial (wall-normal) position of the centre of mass of individual bubbles. Since the bubbles are finite-size and collide with the cylinders there is no data in the PDF very close to $\tilde r=0$ and $\tilde r=1$ which corresponds to the IC and OC, respectively.}
\label{fig:dr-rad}
\end{figure}

We now present results on the influence of the immersed bubbles on the carrier flow dynamics. The net percentage of drag reduction in a two-phase TC flow (here bubbles injected into carrier fluid) can be computed from the kinetic energy transport budget of the system as follows (see Appendix of \citet{sugiyama2008microbubbly} for a detailed derivation):

\begin{equation}
DR = 100 [\langle DR_1 \rangle+\langle DR_2\rangle]=100 \Big [\Big(1-\frac{\langle \epsilon_t\rangle}{\langle \epsilon_s \rangle}\Big) + \Big(\frac{\langle \textbf f_b \cdot \textbf u \rangle}{\langle\epsilon_s\rangle}\Big)\Big]
\label{eqn:dr}
\end{equation}
Here $\langle\epsilon_t\rangle$ and $\langle\epsilon_s\rangle$ are the mean kinetic energy dissipation rate per unit mass of the carrier fluid in the two-phase (with bubbles) and single-phase (without bubbles) cases, respectively; $\textbf f_b$ is the volume averaged source term arising from the effect of the dispersed phase on the carrier fluid (cf. equation \ref{eqn:ns}); $\textbf u$ is the local fluid velocity. A similar decomposition of the stresses transported in a multiphase channel flow is presented in \citet{zhang2010physics}. The two terms on the right hand side of equation (\ref{eqn:dr}), which are named $\langle DR \rangle_1$ and $\langle DR \rangle_2$, correspond to different effects in the flow and the implications of this decomposition is discussed later. In figure \ref{fig:dr-rad}(a), we plot the overall drag reduction $DR$ versus the Weber number ($We$) of the bubbles injected into the flow for both low $Re_i$ and high $Re_i$ cases. One can clearly observe an increasing trend in the drag reduction with increasing Weber number (i.e. increasing bubble deformability).  

\begin{figure}
\centering
\includegraphics[scale=0.75]{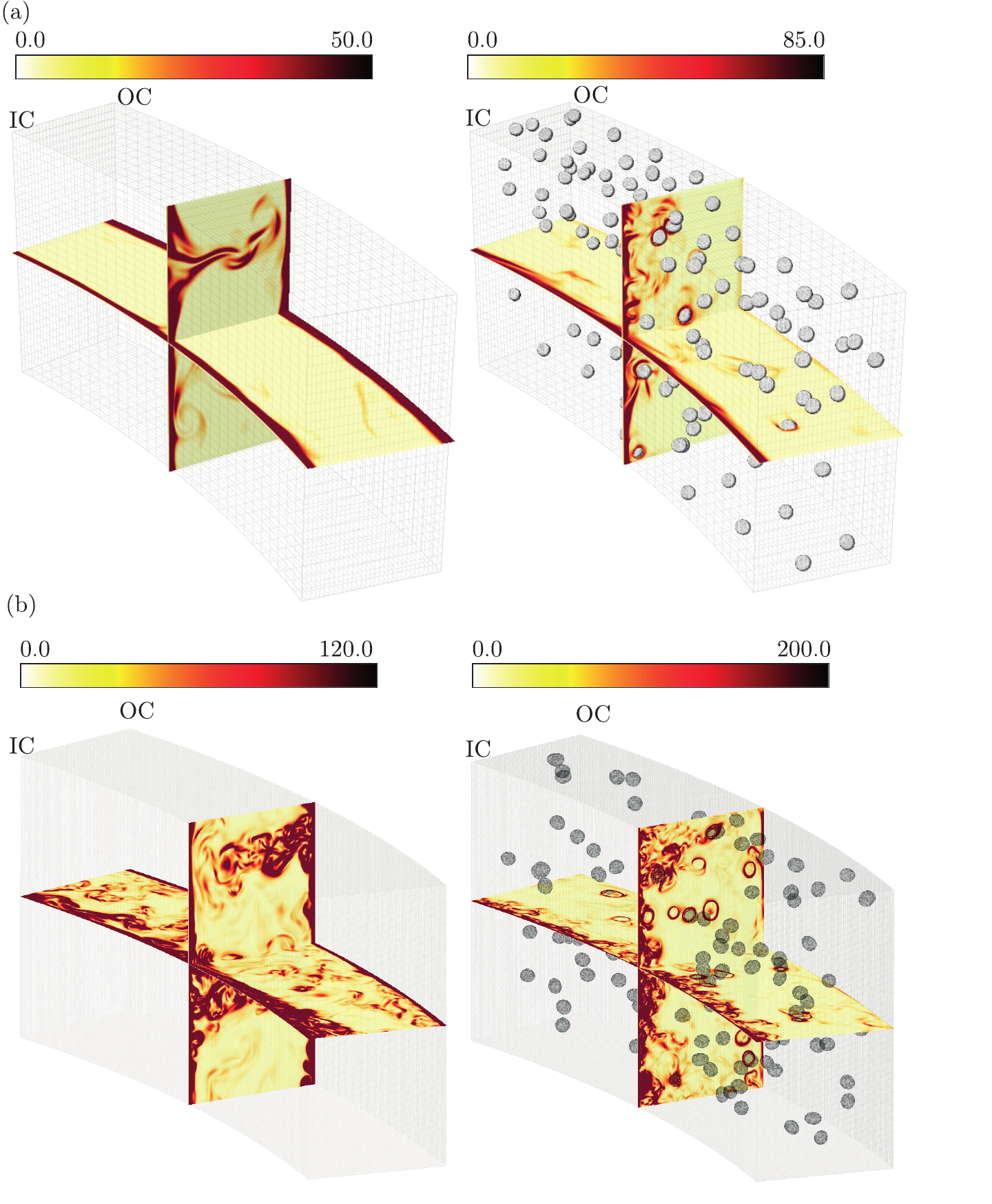}
\caption{Instantaneous snapshot of normalised dissipation $\hat \epsilon=\epsilon/\nu(U/d)^2$ in the mid-azimuthal and mid-axial planes for single phase (left panels) and two-phase flow (right panels). (a) $Re_i=5\times 10^3, We = 1.0$, (b) $Re_i=2\times 10^4, We=1.0$. Dark colour indicates high dissipation and a light colour indicates low dissipation. The dissipation in the bulk and near the OC is attenuated in the presence of bubbles (right panels) as compared to the single phase flow (left panels).}
\label{fig:diss}
\end{figure}

Before analysing the trend in drag reduction with increasing Weber number, it is useful to understand the origin of drag reduction with buoyant bubbles. Similar to the effect seen in \citet{spandan2016drag}, the buoyancy of bubbles causes a mean slip in the direction of the gravity (axial-direction) and in doing so, the bubbles weaken the plumes ejected from the inner cylinder which are primarily wall-normal velocity fluctuations. The weakening of plumes occurs primarily through an axial (direction of buoyancy) perturbation of the wall-normal velocity fluctuations. By weakening the plumes, the bubbles also reduce the effective momentum transport from the inner cylinder to the outer cylinder which consequently leads to drag reduction. 

When the dispersed bubbles are sub-Kolmogorov, deformable bubbles prefer to accumulate near the IC which enhances drag reduction because a higher concentration of bubbles near the IC is more effective in disrupting plume ejections \citep{spandan2017deformable}. In order to understand whether the increase in drag reduction with increasing deformability of finite-size bubbles has any correlation with local bubble accumulation near the IC, we plot the probability distribution function of the normalised radial position $\tilde r=(r-r_i)/(r_o-r_i)$ of the centre of mass of all bubbles in figure \ref{fig:dr-rad}(b). In all cases, the bubbles prefer to position themselves near the IC due to centrifugal forces. However, in contrast to sub-Kolmogorov bubbles, we find no significant difference in the accumulation patterns of finite-size bubbles with increasing deformability. A possible explanation behind this distribution is that finite-size bubbles are less affected by small-scale flow structures ($\eta_K-5 \eta_K$) in comparison to sub-Kolmogorov bubbles. This suggests the presence of additional mechanisms which enhance drag reduction with increasing deformability in the case of finite-size bubbles. 
%It is important to note that similar observations were made in experiments by \citet{van2013importance} (although at much higher Reynolds numbers) where it was shown that a higher local gas concentration near the inner wall does not guarantee stronger drag reduction.

In order to show clearly the origin of drag reduction, in figure \ref{fig:diss}, we show pseudo-colour plots of the normalised kinetic energy dissipation rate in both single phase and two-phase cases. In the case of single phase flow (left panels in figure \ref{fig:diss}), one can clearly observe the large-scale plumes on the IC and OC for both $Re_i$ and that the plumes are noticeably more turbulent in the high $Re_i$ case. The magnitude of dissipation is high around the plumes owing to large strain-rates and also in the boundary layers on the IC and OC. Upon injection of bubbles (right panels in figure \ref{fig:diss}), the plumes originating from the IC no longer protrude into the bulk, but rather become a collection of small-scale intermittent plumes along the surface. Consequently, the turbulence levels in the bulk and near the OC are much weaker as compared to single phase flow. The dispersed bubbles block/weaken the momentum transport such that turbulence is enhanced and confined to regions close to the IC, while it is strongly attenuated in the bulk and near the OC. Remarkably, the weakening of large-scale plumes by the bubbles and subsequent turbulence attenuation in the bulk and OC overpowers the turbulence enhancement near the IC, the combined effect of which leads to a net positive drag reduction. 

\begin{figure}
\centering
\includegraphics[scale=0.75]{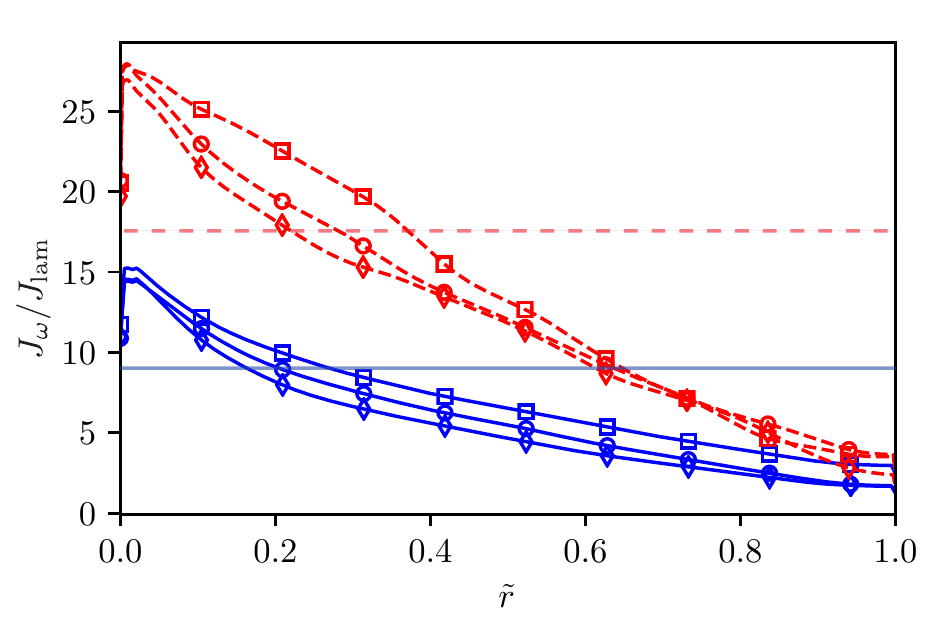}
\caption{Radial profiles of angular velocity transport for $Re_i=5\times 10^3$ (blue solid lines) and $Re_i=2\times 10^4$ (red dashed lines). Squares, circles and diamond markers correspond to $We=0.5, 1, 2$, respectively. Horizontal lines represent the angular velocity flux for the corresponding single phase flow.}
\label{fig:jfl-all}
\end{figure}

To quantify the effect seen in figure \ref{fig:diss}, we look at the angular velocity transport from IC to OC. In single phase flow, the angular velocity flux is conserved in the radial (wall-normal) direction i.e. $\partial_r[r^3\langle u_r\omega\rangle - r^3\nu\partial_r\langle \omega\rangle]=\partial_r[J_\omega]=0$ \citep{eckhardt2007torque}. In two-phase flows, due to the forcing from the dispersed phase, this expression is corrected accordingly and is written as 
\begin{equation}
\partial_r[r^3\langle u_r\omega\rangle - r^3\nu\partial_r\langle \omega\rangle]+r^2f_\theta=\partial_r[J_\omega]+F_\theta=0 
\label{eqn:jomg}
\end{equation}
In these expressions, $J_\omega$ is the angular velocity flux, $\omega$ is the angular velocity, $u_r$ is the radial velocity, $r$ is the wall-normal position, $\langle...\rangle$ refers to averaging in the homogeneous directions (azimuthal and axial) and over time; $f_\theta$ is the forcing from the dispersed phase onto the carrier fluid in the azimuthal (stream-wise) direction. The balance arises naturally from equation \ref{eqn:ns} following the same approach as described in \citet{eckhardt2007torque}, but with the inclusion of a source term in the momentum equation due to the dispersed bubbles. Here, we note that equation \ref{eqn:jomg} is used as the convergence criteria for two-phase simulations. After this condition is satisfied, we collect statistics over approximately 20 full cylinder rotations. To analyse the stresses that are transported from the IC to OC, we plot the azimuthally, axially and time averaged value of $J_\omega$ normalised with its corresponding laminar value $J_{\text{lam}}$ versus the normalised radial position in figure \ref{fig:jfl-all}. 

As mentioned earlier, in single phase flow, $J_\omega$ is constant along the radial position (cf. figure \ref{fig:jfl-all}). In the presence of bubbles, the value $J_\omega$ is higher than that of single phase flow close to the IC and relatively lower in the bulk and near the OC. A lower value of $J_\omega$ in comparison to the single phase case indicates attenuation of turbulence in the flow i.e. the bubbles reduce the radial and axial velocity fluctuations and in turn the stresses that are transported between the cylinders (in the case of purely laminar flow $J_\omega/J_\text{lam}=1$ ). 
Close to the IC, $J_\omega$ is much higher than that of a single phase flow due to two effects. Firstly, collision of bubbles with the IC increases local dissipation in the region between the IC and the bulk. Second, as noted in figure \ref{fig:dr-rad}(b), bubbles tend to prefer regions close to the inner cylinder due to centrifugal forces and since we model the interfaces of bubbles with no-slip boundary conditions (contaminated interface), the fluid in between the inner cylinder and the surface of the bubbles is sheared strongly which leads to an increase in dissipation. The increase in dissipation due to bubble accumulation near the walls is also seen in channel flows injected with bubbles \citep{dabiri2013transition}. Furthermore, the wakes of the bubbles are also regions of intense dissipation which can be observed in figure \ref{fig:diss}. 

\begin{figure}
\centering
\includegraphics[scale=0.75]{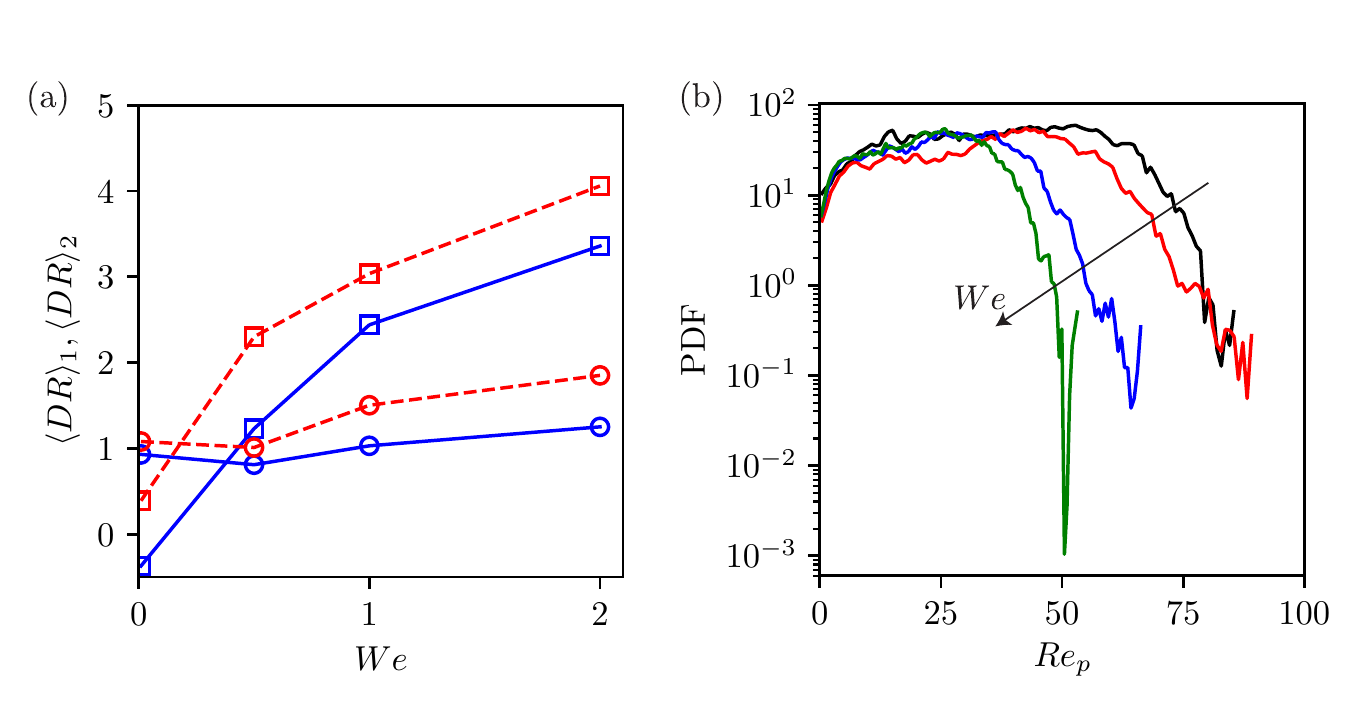}
\caption{(a) Individual contributions from $\langle DR\rangle_1$ (squares) and $\langle DR\rangle_2$ (circles) to $DR$ versus Weber number for $Re_i=5\times 10^3$ (blue solid lines) and $Re_i=2\times10^4$ (red dashed lines). (b) Probability distribution function of the bubble Reynolds number $Re_p$ for different Weber numbers for $Re_i=2\times 10^4$. Arrow indicates increasing Weber number; colour coding of lines similar to figure \ref{fig:dr-rad}}
\label{fig:con-rey}
\end{figure}

We now move on to understanding the role of deformability, i.e. the Weber number, on drag reduction. As shown in equation (\ref{eqn:dr}), the overall drag reduction $DR$ can be written as sum of two terms which correspond to two different physical effects. This formulation of $DR$ is useful in distinguishing the various mechanisms by which a dispersed phase influences the carrier phase and consequently the overall drag in the flow. The first term on the right hand side of equation (\ref{eqn:dr}) is a contribution from the cumulative effect of the reduction in dissipative vortex structures in the flow. If the overall dissipation in the two-phase case $\langle \epsilon_t\rangle$ is less than that of the single-phase case $\langle \epsilon_s\rangle$, $\langle DR_1 \rangle=1-\langle \epsilon_t\rangle/\langle \epsilon_s \rangle$ contributes positively to the overall drag reduction. The second term $\langle DR_2 \rangle$ on the other hand is a contribution from the momentum transfer from the bubbles to the carrier liquid and in a physical sense either assists or impedes the IC in driving the fluid depending on the sign of the term $\langle DR\rangle_2$. In figure \ref{fig:con-rey}(a), we plot the individual contributions from both these terms for the low and high $Re_i$ cases. We find that the contribution from $\langle DR_1 \rangle$ increases significantly with deformation in comparison to $\langle DR_2 \rangle$. The energy injection rate $\langle DR\rangle_2$ does not change significantly in comparison with $\langle DR\rangle_1$. This is expected as due to the averaging in $\langle DR\rangle_2$ in both space and time, the primary source of momentum transfer between the bubbles and the carrier fluid comes from the buoyancy of the bubbles which does not change with deformability. In all cases, we ensure that every individual bubble has the same volume and is incompressible. It is also important to note here that in contrast to sub-Kolmogorov bubbles, because of the strong buoyancy of an individual finite-size bubble, they rarely get trapped in the descending part of the Taylor roll which may lead to a negative $\langle DR\rangle_2$.

\begin{figure}
\centering
\includegraphics[scale=0.5]{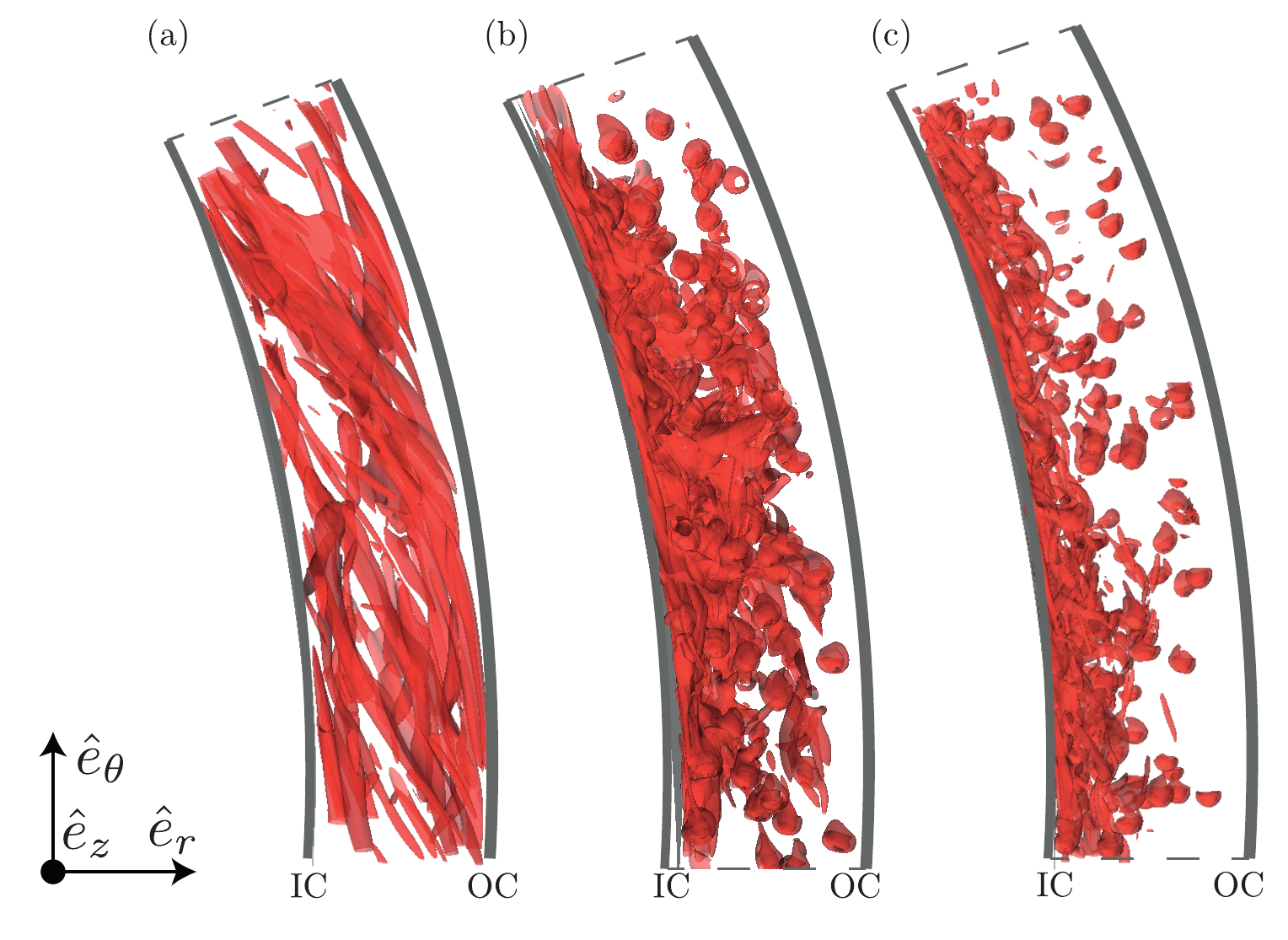}
\caption{Vortical structures mapped using the Q-criterion for $Re_i=5\times 10^3$ (a) Single phase (b) Two-phase $We\sim0$ (c) Two-phase $We\sim2.0$ }
\label{fig:qstruc}
\end{figure}

While the term $\langle DR\rangle_1$ is overall positive for all cases, it is negative only at low $Re_i$ with rigid bubbles. In this case, the dissipation in the wake generated by the rigid bubbles overcomes the dissipation in large-scale coherent structures. It is difficult to generalise this effect given that $d_b/\eta_K$ is different for the low and high $Re_i$ cases. To understand the primary source of increasing drag reduction with increasing deformability, we must understand the trend of $\langle DR\rangle_1$ and for this we look into the effect of finite-size bubbles on the surrounding turbulent flow. 
Finite-size bubbles with sufficiently high slip velocity and particle (local) Reynolds number shed wakes which are additional sources of dissipation in the flow. While bubble wakes typically increase dissipation, the bubbles themselves also weaken the large scale plumes which are the primary sources for momentum transport (cf. figure \ref{fig:diss} and \ref{fig:jfl-all}); the weakening of plumes overcomes the dissipation induced by bubbles wakes thus leading to a net positive drag reduction. When the immersed bubbles are made more deformable by lowering their surface tension (or increasing the Weber number), they are stretched along the stream-wise direction similar to that of sub-Kolmogorov deformable bubbles \citep{spandan2017deformable}. The stretching leads to a lower projected surface area in the direction of the relative velocity which in turn leads to a lower bubble Reynolds number. This is shown in figure \ref{fig:con-rey}(b), where we plot the probability distribution function of the bubble Reynolds number defined as $Re_p=(|\textbf u-\textbf v|)\sqrt{4\bar A/\pi}/\nu$ for $Re_i=2\times 10^4$. Here it is important to note that the exact calculation of particle Reynolds number in simulations involving IBM is non-trivial as the local fluid velocity $\textbf u$ in $Re_p$ is ill-defined. A discussion on the calculation is given in Appendix \ref{appA}. 

It is easily seen that the bubble Reynolds number distribution is much wider for rigid non-deformable ($We=0.01$) bubbles in comparison to deformable ($We=2.0$) bubbles. We have found no significant differences in the distribution of the relative velocities between individual bubbles and the local fluid velocity $|\textbf u-\textbf v|$; the differences in the projected areas of the bubbles ($\bar A$) is the primary source of differences in the distribution of $Re_p$. The Reynolds numbers of deformable bubbles are much lower in comparison to that of rigid non-deformable bubbles, which corresponds to a relatively lower dissipation induced in the flow by the bubble wake. Thus, while the energy injection rate from the bubbles remains roughly the same for bubbles with varying Weber numbers, the dissipation originating from the bubbles' wakes decreases with increasing deformability. This leads to a positive contribution to the drag reduction from the term $\langle DR\rangle_1$. When the bubbles are non-deformable, the positive contribution to $DR$ induced by the momentum forcing term $\langle DR\rangle_2$ is compensated by the dissipation arising from the wake of non-deformable bubbles as can be seen for $We=0.01$ in figure \ref{fig:con-rey}(a). This is shown qualitatively by visualising iso-contours of vortical structures through the Q-criterion in figure \ref{fig:qstruc}. While one can clearly see the coherent Taylor rolls for the single phase case, with the inclusion of bubbles, dissipative vortices change completely with a high concentration near the inner cylinder. In the case of high Weber number, the dissipative structures are much weaker as compared to that of the flow with rigid bubbles.

\section{Summary and Outlook}
\label{sec:sum}
With the help of fully resolved numerical simulations, we have been able to identify different physical effects and their individual contributions to bubble induced drag reduction in turbulent Taylor-Couette (TC) flow. We show that drag reduction in two-phase TC flow is an effect of the bubbles weakening the large-scale plumes responsible for momentum transport. The net effect of immersed bubbles is that in comparison to single phase flow, turbulence near the inner cylinder is enhanced (due to dissipation in between the bubbles and the bounding wall, from bubble wake and bubble collision with walls) while turbulence is attenuated in the bulk and near the outer cylinder (due to bubbles blocking the momentum transfer between the cylinders). Furthermore, we have identified different physical mechanisms and their corresponding contributions to drag reduction. The net drag reduction is written as a sum of two terms which are contributions from (i) changes in dissipative structures in the flow and (ii) a energy injection rate from the dispersed bubbles. 
We find that the intensity of dissipative structures in the flow decreases with increasing deformability. Due to stretching in the stream-wise direction, deformable bubbles have on average a lower local Reynolds number in comparison to non-deformable bubbles which leads to reduced dissipation in the flow. While the energy injection rate from the bubbles always contributes to drag reduction, the net effect from dissipation becomes the governing factor in achieving significant drag reduction. 

The current numerical simulations only operate in the regime where coalescence and breakup of bubbles is avoided which ensures a mono-dispersed bubble distribution. However, in a real system, coalescence and breakup of bubbles might lead to a distribution of sizes in which case their effect on the underlying turbulent structures may vary and we are currently working on implementing these features within our numerical framework.  

This work was supported by the Netherlands Center for Multiscale Catalytic Energy Conversion (MCEC), an NWO Gravitation programme funded by the Ministry of Education, Culture and Science of the government of the Netherlands. The simulations were carried out on the national e-infrastructure of SURFsara, a subsidiary of SURF cooperation, the collaborative ICT organization for Dutch education and research. We also acknowledge PRACE for awarding us access to Marconi super computer, based in Italy at CINECA under PRACE project number 2016143351.

\appendix
\section{}\label{appA}
In numerical simulations involving fully resolved techniques (such as IBM), calculating $Re_p$ is extremely non-trivial due to two primary reasons. First, while the mean bubble velocity ($\textbf v$) is clearly defined and is computed in the simulation, the mean fluid velocity ($\textbf u$) that the bubble experiences needs to be computed through a filtering approach. In this paper, we compute the mean of the fluid velocity interpolated at the end of a probe from each Lagrangian marker (see \citet{spandan2017parallel} for more details on the probe technique). Second, the calculation of a mean projected area of a deformable body along specific directions is done using a coarse-grained ray-tracing technique. This is achieved by computing the projected areas of individual Eulerian cells immersed inside the bubble along specific directions and the mean projected area is then computed as $\bar A=\sum_{i=1}^{i=3} \gamma_i A_i$, where $\gamma_i$ are the direction cosines of the relative velocity vector and $A_i$ is the projected area in the $i^{\text th}$ direction.

\bibliographystyle{jfm}
%% Note the spaces between the initials
\bibliography{../../mylit}

\end{document}